\documentclass[fleqn,12pt,twoside]{article}
\usepackage{graphicx,amsmath}
\usepackage[german]{babel}
\usepackage{espcrc1}

\begin{document}

\author{Ph. H\"{a}gler\address{University of
Regensburg, D-93040 Regensburg,  Germany}
, B. Pire\address{CPhT, {\'E}cole Polytechnique, 
F-91128 Palaiseau,
France}, 
L. Szymanowski$^{\text{b},}$\address{So{\l}tan Institute for Nuclear Studies,
Ho\.za 69, 00-681 Warsaw, 
Poland} and 
O.V. Teryaev\address{Bogoliubov Lab. of Theoretical Physics
, JINR, 141980 Dubna
, Russia}}
\title{The charge asymmetry from Pomeron-Odderon interference in hard diffractive $%
\pi ^{+}\pi ^{-}$-electroproduction\thanks{%
Talk presented by Ph.H. based on the joint publication \protect\cite{HPST}.}}
\maketitle

\begin{abstract}
The interference of Pomeron and Odderon amplitudes gives rise to a charge
asymmetry in the diffractive electroproduction of a $\pi ^{+}\pi ^{-}$%
-pair. We calculate this charge asymmetry in perturbative QCD (pQCD) in
the Born approximation and on the leading $Q^{2}$-level. The numerical
evaluation shows a sizeable asymmetry in an experimentally accessible
kinematical region. We find a characteristic $m_{\pi ^{+}\pi ^{-}}$%
-dependence mainly dictated by the relevant Breit-Wigner-amplitudes and
the corresponding phase-shifts.
\end{abstract}

\vspace*{0.2cm}
Pomeron and Odderon exchanges in pQCD are to lowest
order given by colour singlet exchanges in the $t$-channel with two and
three gluons, respectively. While the Pomeron is described by the well known
solutions of the Balitsky-Fadin-Kuraev-Lipatov (BFKL) equation \cite{BFKL},
only in recent years progress in solving the corresponding
Bartels-Kwiecinski-Praszalowicz (BKP) equation \cite{BKP} for the
Odderon has been reported (see \cite{BLV} and \cite{DKKM}).

Although the Pomeron and the Odderon exchange both theoretically induce
dominant contributions to high energy hadronic cross sections, so far only
the Pomeron exchange has been confirmed in the comparison of theory and
experiment. Pure Odderon exchange, which e.g. has been considered in the
case of diffractive $\eta _{c}$-meson photo- and electroproduction in the
Born approximation in QCD, introduces rather small cross sections \cite
{KM,Engel}. However, the status of Odderon induced reactions is not
settled. The inclusion of evolution following from the BKP equation \cite
{Vacca2} leads to an increase of the predicted $\eta _{c}$-meson
photoproduction cross section by one order of magnitude. Unfortunately
recent experimental studies at HERA of exclusive $\pi ^{0}$ photoproduction 
\cite{Olsson} point towards a very small cross section, which confusingly
stays in disagreement with theoretical predictions based on the stochastic
vacuum model \cite{Dosh}.

In order to push the hunt for the Odderon forward, it seems therefore to be
highly valuable to study Odderon effects at the amplitude level by means of
asymmetries, utilising the different (even and odd) charge conjugation
properties of the Pomeron and the Odderon. Such asymmetries $\mathcal{A}$
are approximately proportional to the ratio of the amplitude of the Odderon
induced reaction $\mathcal{M}_{\mathcal{O}}$ and the corresponding amplitude
for the Pomeron exchange $\mathcal{M}_{\mathcal{P}}$, $\mathcal{A}\tilde{%
\propto}\left| \mathcal{M}_{\mathcal{O}}\right| /\left| \mathcal{M}_{%
\mathcal{P}}\right| $, in contrast to seemingly much smaller cross-section
ratios $\mathcal{R}$ in pure Odderon mediated reactions which would behave
like $\mathcal{R}\tilde{\propto}\left| \mathcal{M}_{\mathcal{O}}\right|
^{2}/\left| \mathcal{M}_{\mathcal{P}}\right| ^{2}.$

Brodsky et al. \cite{Brodsky} suggested in this context to study the Odderon
by the use of asymmetries in open charm production. Since the final state
charm-anticharm system has no definite charge parity both Pomeron and
Odderon exchanges contribute to this process, and the asymmetry allows to
project out the pure interference contribution. Similarly Ivanov et al. \cite
{Nikolaev} studied recently the charge asymmetry in the soft photoproduction
of a $\pi ^{+}\pi ^{-}$-pair. In both approaches (\cite{Brodsky} and \cite
{Nikolaev}) the Pomeron and the Odderon are treated
non-perturbatively.

\begin{figure}[t]
\centerline{\includegraphics[scale=.5]{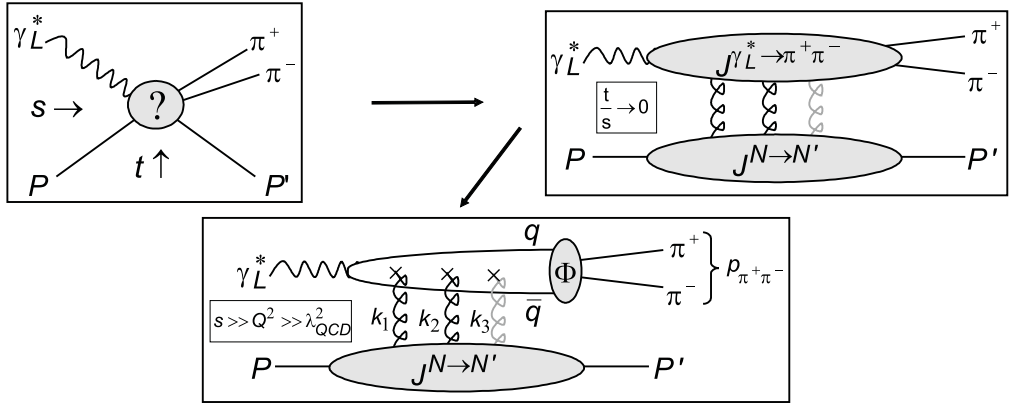}}
\caption{The process $\gamma ^{*}+P\rightarrow \pi
^{+}\pi ^{-}+P.$} 
\label{factorization}
\end{figure}

\begin{figure}[t]
\begin{minipage}[t]{80mm}
\centerline{\includegraphics[scale=.52]{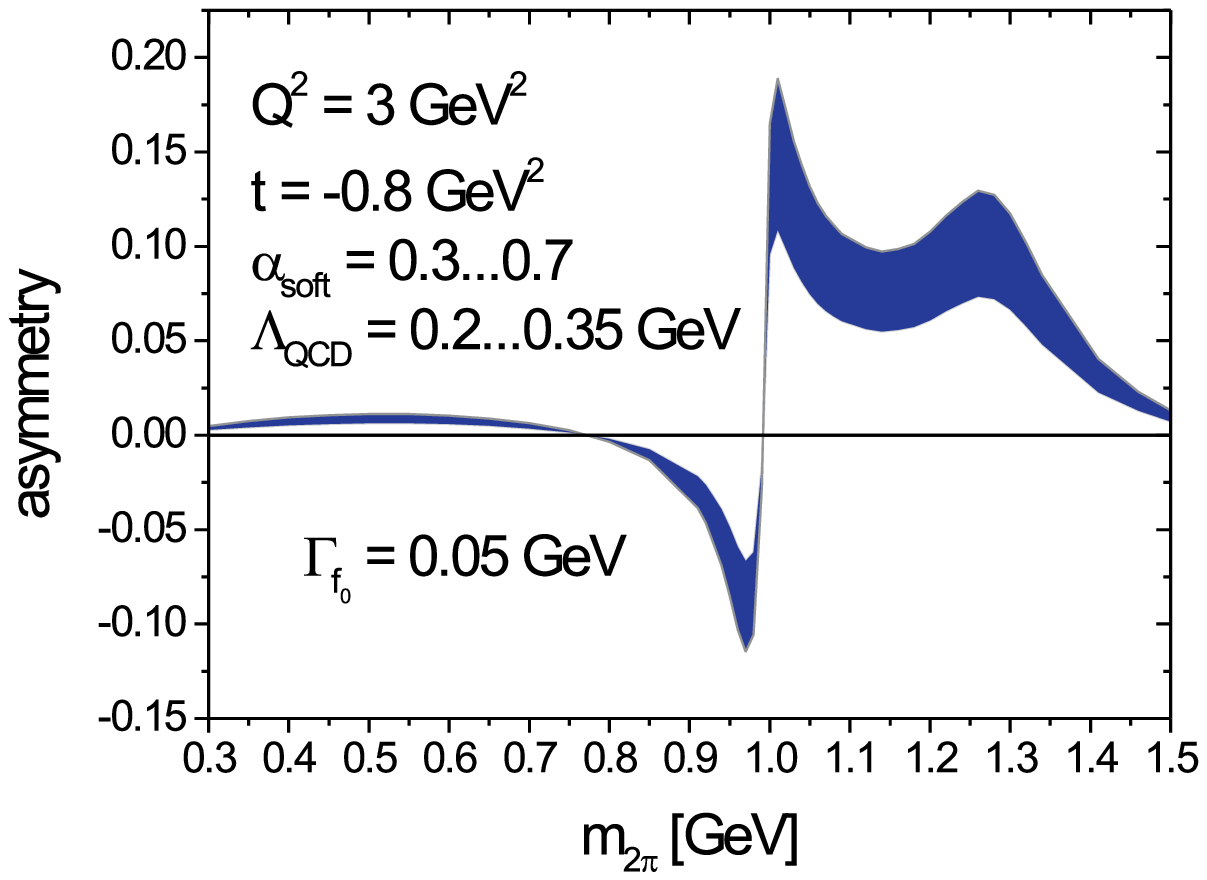}}
\caption{The $m_{2\pi}$-dependence.}
\label{asym1}
\end{minipage}
\hspace{\fill}
\begin{minipage}[t]{80mm}
\centerline{\includegraphics[scale=.52]{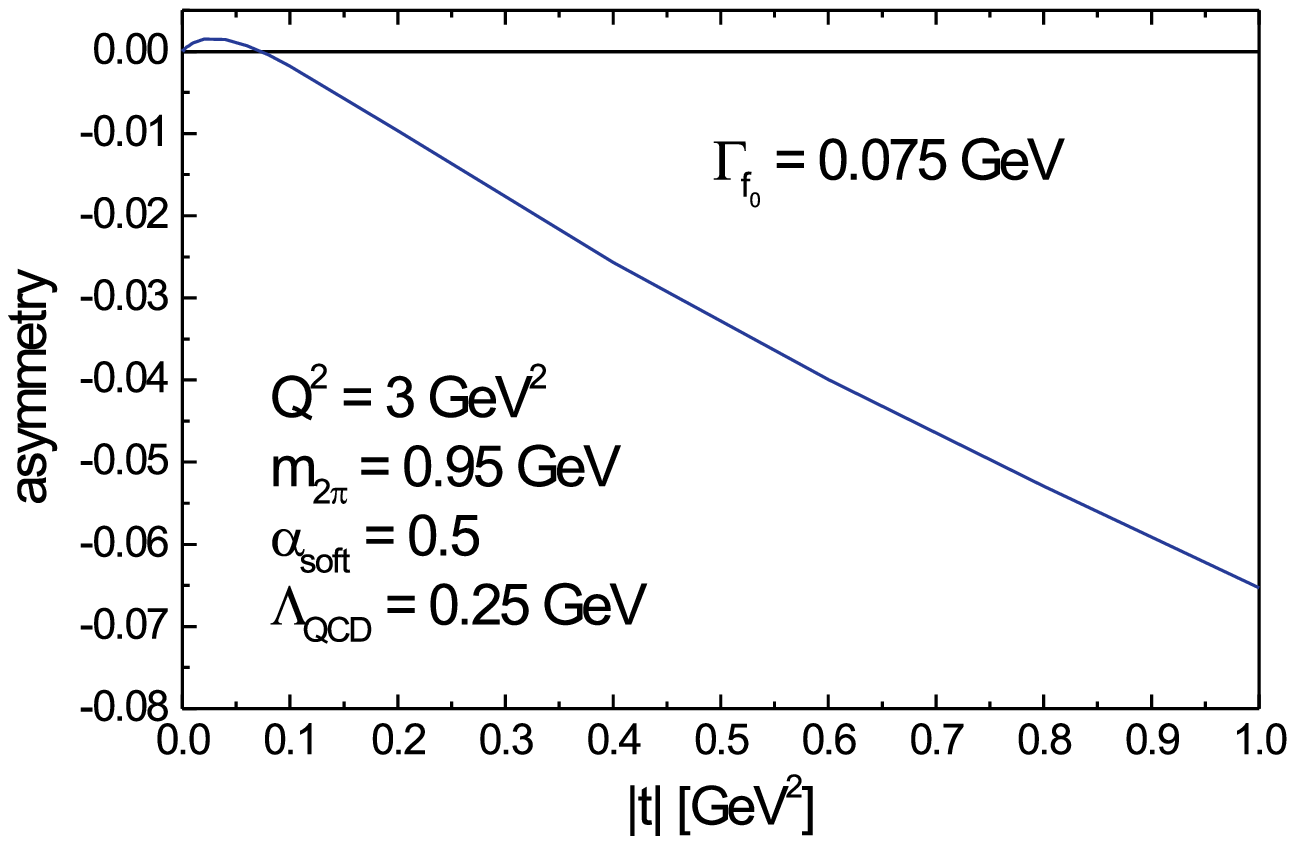}}
\caption{ The $t$-dependence.
}
\label{asym2}
\end{minipage}
\vspace*{-.4cm}
\end{figure}

We carry over the idea of Nikolaev et al. to the case of the charge
asymmetry in diffractive $\pi ^{+}\pi ^{-}$-electroproduction to search for
the QCD-Odderon at the amplitude level. In contrast to the previous works we
investigate the process $\gamma ^{*}+P\rightarrow \pi ^{+}\pi ^{-}+P$ at
high energies squared $s$ but small momentum transfer squared $t$ and large
photon virtuality squared $Q^{2}$. This permits us in the first place to
apply the $k_{\bot }$-factorization, secondly to adopt the
collinear approximation for the subprocess $q\bar{q}\rightarrow \pi ^{+}\pi
^{-}$ (with $q=u,d$) and thirdly to
treat the Pomeron and the Odderon as two respectively three gluon exchanges
within QCD perturbation theory. Since we limit ourself to leading contributions
in $Q^{2}$, only a longitudinal virtual photon is  considered. A
scheme of our approximations is shown in fig. \ref{factorization}.
The above explained approach allows us to write the corresponding amplitudes
for the Pomeron and Odderon exchange in the following
impact-parameter-representation
\begin{equation}
\mathcal{M}_{\mathcal{P}}=-i\,s\,\int \;\frac{d^{2}\vec{k}_{1}\;d^{2}\vec{k}%
_{2}\;\delta ^{(2)}(\vec{k}_{1}+\vec{k}_{2}-\vec{p}_{\pi ^{+}\pi ^{-}})}{%
(2\pi )^{2}\,\vec{k}_{1}^{2}\,\vec{k}_{2}^{2}}J_{\mathcal{P}}^{\gamma
^{*}\rightarrow \pi ^{+}\pi ^{-}}(\vec{k}_{1},\vec{k}_{2})\cdot J_{\mathcal{P%
}}^{N\rightarrow N^{\prime }}(\vec{k}_{1},\vec{k}_{2}),  \label{pom}
\end{equation}
\begin{equation}
\mathcal{M}_{\mathcal{O}}=-\frac{8\,\pi ^{2}\,s}{3!}\int \;\frac{d^{2}\vec{k}%
_{1}\;d^{2}\vec{k}_{2}d^{2}\vec{k}_{3}\;\delta ^{(2)}(\vec{k}_{1}+\vec{k}%
_{2}+\vec{k}_{3}-\vec{p}_{\pi ^{+}\pi ^{-}})}{(2\pi )^{6}\,\vec{k}_{1}^{2}\,%
\vec{k}_{2}^{2}\,\vec{k}_{3}^{2}}J_{\mathcal{O}}^{\gamma ^{*}\rightarrow \pi
^{+}\pi ^{-}}\cdot J_{\mathcal{O}}^{N\rightarrow N^{\prime }}.  \label{odd}
\end{equation}
All variables are explained in fig.\ref{factorization}. The impact factors $%
J_{\mathcal{P},\mathcal{O}}^{\gamma ^{*}\rightarrow \pi ^{+}\pi ^{-}}$ can
in part be
perturbatively calculated and are given by 
\begin{eqnarray}
J_\mathcal{P}^{\gamma _{L}^{*}\rightarrow \pi ^{+}\pi ^{-}}(\vec{k}_{1},\vec{k}_{2})
&=&-\frac{i\,e\,g^{2}\,\delta ^{ab}\,Q}{2\,N_{C}}\;\int\limits_{0}^{1}\,dz\,z%
{\bar{z}}\,P_\mathcal{P}(\vec{k}_{1},\vec{k}_{2})\,\Phi ^{I=1}(z,\zeta ,m_{\pi
^{+}\pi ^{-}}^{2}),  \label{lP} \\\nonumber
P_\mathcal{P}(\vec{k}_{1},\vec{k}_{2}) &=&\frac{1}{z^{2}\vec{p}_{\pi ^{+}\pi
^{-}}^{2}+\mu ^{2}}+\frac{1}{{\bar{z}}^{2}\vec{p}_{\pi ^{+}\pi ^{-}}^{2}+\mu
^{2}}-\frac{1}{(\vec{k}_{1}-z\vec{p}_{\pi ^{+}\pi ^{-}})^{2}+\mu ^{2}} \\\nonumber
&&-\frac{1}{(\vec{k}_{1}-{\bar{z}}\vec{p}_{\pi ^{+}\pi ^{-}})^{2}+\mu ^{2}}
\end{eqnarray}
for the Pomeron and 
\begin{eqnarray}
 J_\mathcal{O}^{\gamma _{L}^{*}\rightarrow \pi ^{+}\pi ^{-}}(\vec{k}_{1},\vec{k}%
_{2},\vec{k}_{3})&=&-\frac{i\,e\,g^{3}\,d^{abc}\,Q}{4\,N_{C}}%
\;\int\limits_{0}^{1}\,dz\,z{\bar{z}}\,P_\mathcal{O}(\vec{k}_{1},\vec{k}_{2},\vec{k}%
_{3})\,\frac{1}{3}\Phi ^{I=0}(z,\zeta ,m_{\pi ^{+}\pi ^{-}}^{2}), \\\nonumber
 P_\mathcal{O}(\vec{k}_{1},\vec{k}_{2},\vec{k}_{3})&=&\frac{1}{z^{2}\vec{p}_{\pi
^{+}\pi ^{-}}^{2}+\mu ^{2}}-\frac{1}{{\bar{z}}^{2}\vec{p}_{\pi ^{+}\pi
^{-}}^{2}+\mu ^{2}} \\\nonumber
 &&-\sum\limits_{i=1}^{3}\left( \frac{1}{(\vec{k}_{i}-z\vec{p}_{\pi ^{+}\pi
^{-}})^{2}+\mu ^{2}}-\frac{1}{(\vec{k}_{i}-{\bar{z}}\vec{p}_{\pi ^{+}\pi
^{-}})^{2}+\mu ^{2}}\right) 
\end{eqnarray}
for the Odderon.
The variable $\mu $ is
defined by $\mu ^{2}=m_{q}^{2}+z\,{\bar{z}}\,Q^{2}$, where ${\bar{z}\equiv
1-z}$, $m_{q}$ is
the quark mass, and we put $m_{u}\simeq m_{d}$. The 2$\pi $GDAs $\Phi
^{I}(z,\zeta ,m_{\pi ^{+}\pi ^{-}}^{2})$ (see
e.g. \cite{DGPT}-\cite{DGP}) for
isospin $I=0,1$ depend on the longitudinal momentum fraction $z$, the
invariant mass $m_{\pi ^{+}\pi ^{-}}$ of the $\pi ^{+}\pi ^{-}$-system and the variable $%
\zeta $, which is directly related to the polar decay angle $\theta $ of the 
$\pi ^{+}$ in the dipion rest frame by $\beta \cos \theta =2\zeta -1$
and $\beta \equiv \sqrt{1-4m_{\pi}^{2}/m_{\pi ^{+}\pi ^{-}}^{2}}$.
The GDAs have to be modelled and in our case are proportional to the $%
s,p$ and $d$ wave phase factors times the corresponding absolute values of
the Breit-Wigner-amplitudes. The nucleon impact factors $J_{\mathcal{P},%
\mathcal{O}}^{N\rightarrow N^{\prime }}$ likewise cannot be rigorously
calculated. The details of our choices for the functions $\Phi ^{I}(z,\zeta
,m_{\pi ^{+}\pi ^{-}}^{2})$  and 
$J_{\mathcal{P},\mathcal{O}}^{N\rightarrow N^{\prime }}$ can be found in 
\cite{HPST} and the references therein. The charge asymmetry is finaly
defined by \cite{Nikolaev} 
\begin{equation}
\nonumber
\hspace{-0.7cm} \mathcal{A}(Q^{2},t,m_{\pi ^{+}\pi ^{-}}^{2})=\frac{\int \cos \theta
\,d\sigma (s,Q^{2},t,m_{\pi ^{+}\pi ^{-}}^{2},\theta )}{\int d\sigma
(s,Q^{2},t,m_{\pi ^{+}\pi ^{-}}^{2},\theta )}=\frac{\int\limits_{-1}^{1}\,%
\cos \theta \,d\cos \theta \,2\text{Re}\left[ \mathcal{M}_{\mathcal{P}%
}^{\gamma _{L}^{*}}(\mathcal{M}_{\mathcal{O}}^{\gamma _{L}^{*}})^{*}\right] 
}{\int\limits_{-1}^{1}\,d\cos \theta \left[ |\mathcal{M}_{\mathcal{P}%
}^{\gamma _{L}^{*}}|^{2}+|\mathcal{M}_{\mathcal{O}}^{\gamma
_{L}^{*}}|^{2}\right] }.
\end{equation}
Numerical results for the 
$m_{\pi ^{+}\pi ^{-}}$-dependence of the charge asymmetry 
are shown in fig.(\ref{asym1}).
The error-band results
from a combined variation of $\Lambda _{QCD}$ and the nonperturbative
coupling $\alpha _{soft}$ (representing the coupling of the gluons in the
nucleon impact factor). The asymmetry is sizeable in the vicinity of the $%
f_{0}$ and the $f_{2}$-resonance. The $t$-dependent plot in fig.(\ref{asym2}%
) shows a characteristic zero around $\left| t\right| =0.1$ GeV$^{2}$, which
has been already discussed in Ref. \cite{Vacca2}.

We learnt at this conference that  preliminary results
for the charge asymmetry 
are available by now from
the HERMES collaboration \footnote{Thanks to N.Bianchi, private communication}.
Although the interference process at work at medium values
of $x_{Bj}$ 
is 
different from our case \cite{LD}, it is still
quite surprising that there is no sign of the $f_0$-resonance in the
preliminary data. 

We plan to extent the present study to include transversely polarized
photons for the charge asymmetry and to perform calculations of the
single spin asymmetry induced by the
Pomeron-Odderon-interference \cite{HPST2}.

\end{document}